\documentclass[runningheads,a4paper]{llncs}

\usepackage{amssymb}
\usepackage{graphicx}
\usepackage{multirow}
\usepackage{leqno}
\usepackage{url}
\usepackage{subfigure}
\usepackage{amssymb}
\usepackage{url}
\usepackage{float}
\usepackage{epstopdf}
\usepackage{epsfig}

\urldef{\mailsa}\path|{yasmin, aalsum, mweigle, mln}@cs.odu.edu|
\newcommand{\keywords}[1]{\par\addvspace\baselineskip
\noindent\keywordname\enspace\ignorespaces#1}

\begin{document}

\mainmatter  % start of an individual contribution

\title{Who and What Links to the Internet Archive}

%\titlerunning{Lecture Notes in Computer Science: Authors' Instructions}

\author{Yasmin AlNoamany \and Ahmed AlSum \and Michele C. Weigle \and Michael L. Nelson}
\authorrunning{AlNoamany et. al. }

\institute{Old Dominion University, Department of Computer Science \\
Norfolk VA 23529, USA\\
\mailsa}


\toctitle{Lecture Notes in Computer Science}
\tocauthor{Authors' Instructions}
\maketitle

\begin{abstract}
%Internet Archive is the largest and oldest web archive that holds a significant repository of our recent history and cultural heritage. Because of the potential importance of the Internet Archive's Wayback Machine, which provides web archive users with access to a resource of immense value, this study introduces an analysis of the Wayback Machine access logs to determine what web archive users look for, why they come to IA, where they come from, and how pages link to IA.

The Internet Archive's (IA) Wayback Machine is the largest and oldest public web archive and has become a significant repository of our recent history and cultural heritage. Despite its importance, there has been little research about how it is discovered and used.  Based on web access logs, we analyze what users are looking for, why they come to IA, where they come from, and how pages link to IA. We find that users request English pages the most, followed by the European languages. Most human users come to web archives because they do not find the requested pages on the live web. About 65\% of the requested archived pages no longer exist on the live web. We find that more than 82\% of human sessions connect to the Wayback Machine via referrals from other web sites, while only 15\% of robots have referrers. Most of the links (86\%) from websites are to individual archived pages at specific points in time, and of those 83\% no longer exist on the live web.

%Of the referrers, 86\% link to a page at a specific point in time, mostly because the archived pages do not exist on the live web.
\end{abstract}

\section{Introduction}
A variety of research has been conducted for studying web archives in order to answer questions related to user needs and to present web archive data to users \cite{Padia2012,costa2010}. However, no previous work has been carried out to answer these questions: What content languages are web archive users looking for? Why do users come to web archives? Where do web archive users come from? Who links to web archives? How do sites link to web archives? Do sites link deeply to specific archived pages or link to the repository? Why do sites link to the past? 
%What is the distribution of the top languages used in web archives? Where do web archive users come from? Who links to web archives? How do sites link to web archives? Do they link deeply to specific archived pages or link to the whole repository? Why do they link to the past? 

The Internet Archive \cite{wayback:billion} is the first web archiving initiative attempting global scope and currently holds over 240 billion web pages with archives as far back as 1996 \cite{Kahle2013}. It allows traveling back in time for traversing archived versions of web pages through the Wayback Machine \cite{Tofel2007}. This paper provides a study of the requests of web archive users, both humans and robots, to gain insight into what users look for, in the context of the language of the requested pages, through an analysis of the server logs of the Internet Archives' Wayback Machine. We also provide an analysis of referring pages of human users to investigate how humans discover the Wayback Machine, why the referrers link to web archives, and how they link to web archives.

We found that users of Internet Archive's Wayback Machine request English pages the most, followed by several European languages. We also found that most human users come to the Wayback Machine via links or direct address presumably because they did not find the requested pages on the live web. Of the requested archived pages, 65\% do not currently exist on the live web. From analyzing the referrers, we found that more than 82\% of human sessions have referrers, while only 15\% of robot sessions have referrers. We also found that 86\% of the referrers are deep links to archived pages.
%For humans, the majority of the external pages link to the Wayback Machine because they did not find the web pages on the live web.
%In particular, we find that English is the most used language on web archives, then come the European languages which represent, for the successful requests (requests with HTTP 200), 13\%  for humans and 19 for robots. While for the missing web pages that got HTTP 404 status code, they represent 22\% for humans and 23\% for robots. We also find that most human users come to web archives because they do not find the requested pages on the live web. The percentage of the requested archived pages which are not exist in the live web is 65\%. From analyzing the referrers of humans, we find that more than 82\% of human sessions connect to web archives from web sites, while only 15\% of robots sessions have referrers. For humans, 83\% of the referee do not exist on the live web now, which means most of the referrers link to web archives because they do not find the web pages on the live web. Of the referrers, there are 86\% of web sites link to web archive deeply, which means they link to web page in specific time.
%83.04020101

\section{Related Work}
To the best of our knowledge, no prior study has analyzed where web archive users come from nor what they look for in terms of the linguistic context. Furthermore, the usage of web archives in general has not been widely studied. The characterization of search behavior and the information needs of web archive users have been studied by Costa et al.\@ \cite{Costa2011,costa2010} based on quantitative analysis of the Portuguese Web Archive (PWA) search logs. 
 In a previous study \cite{AlNoamany2013}, we provided the first analysis of user access to a large web archive. We discovered four basic access patterns for web archives through analysis of web server logs from the Internet Archive's Wayback Machine. In the study, we applied heuristics for robot detection after data filtering and found that robot sessions outnumber human sessions 10:1. Robots outnumber humans in terms of raw, unfiltered requests 5:4, and 4:1 in terms of megabytes transferred. 

Many studies have investigated what is missing from digital libraries and web archives, in addition to the effect of this on the satisfaction of users' needs and expectations \cite{Thelwall2004,Carmel2008,Zhuang2005,Silva2009}. In \cite{Thelwall2004}, the Internet Archive's coverage of the web was investigated. The results showed an unintentional international bias through uneven representation of different countries in the archive. Carmel et al.\@ \cite{Carmel2008} suggest a tool to dynamically analyze the query logs of the digital library system, identify the missing content queries, and then direct the system to obtain the missing data. 
We investigate what is missing through an analysis of requests with an HTTP 404 status in the Wayback Machine web server logs.

\subsection*{Memento Terminology}
In this section, we explain the terminology we adopt in the rest of the paper. Memento \cite{nelson:memento:tr} is an HTTP protocol extension which enables time travel on the web by linking the current resources with their prior state. Memento defines the following terms:
\begin{itemize}
 \item URI-R identifies the original resource. It is the resource as it used to appear on the live web. A URI-R may have 0 or more mementos (URI-Ms).
\item URI-M identifies an archived snapshot of the URI-R at a specific datetime, which is called Memento-Datetime, e.g., URI-M$_{i}$= URI-R$ @ t_{i}$.
\item URI-T identifies a TimeMap, a resource that provides a list of mementos (URI-Ms) for a URI-R with their Memento-Datetimes, e.g., $URI-T(URI-R) = \{URI-M_{1}, URI-M_{2}, ..., URI-M_{n}\}$.
 \end{itemize}
 Although we use Memento terminology, the logs we analyze are from the Internet Archive's Wayback Machine and not the Memento API.

\begin{figure}[h!]
	\centering
	\includegraphics[scale=0.43]{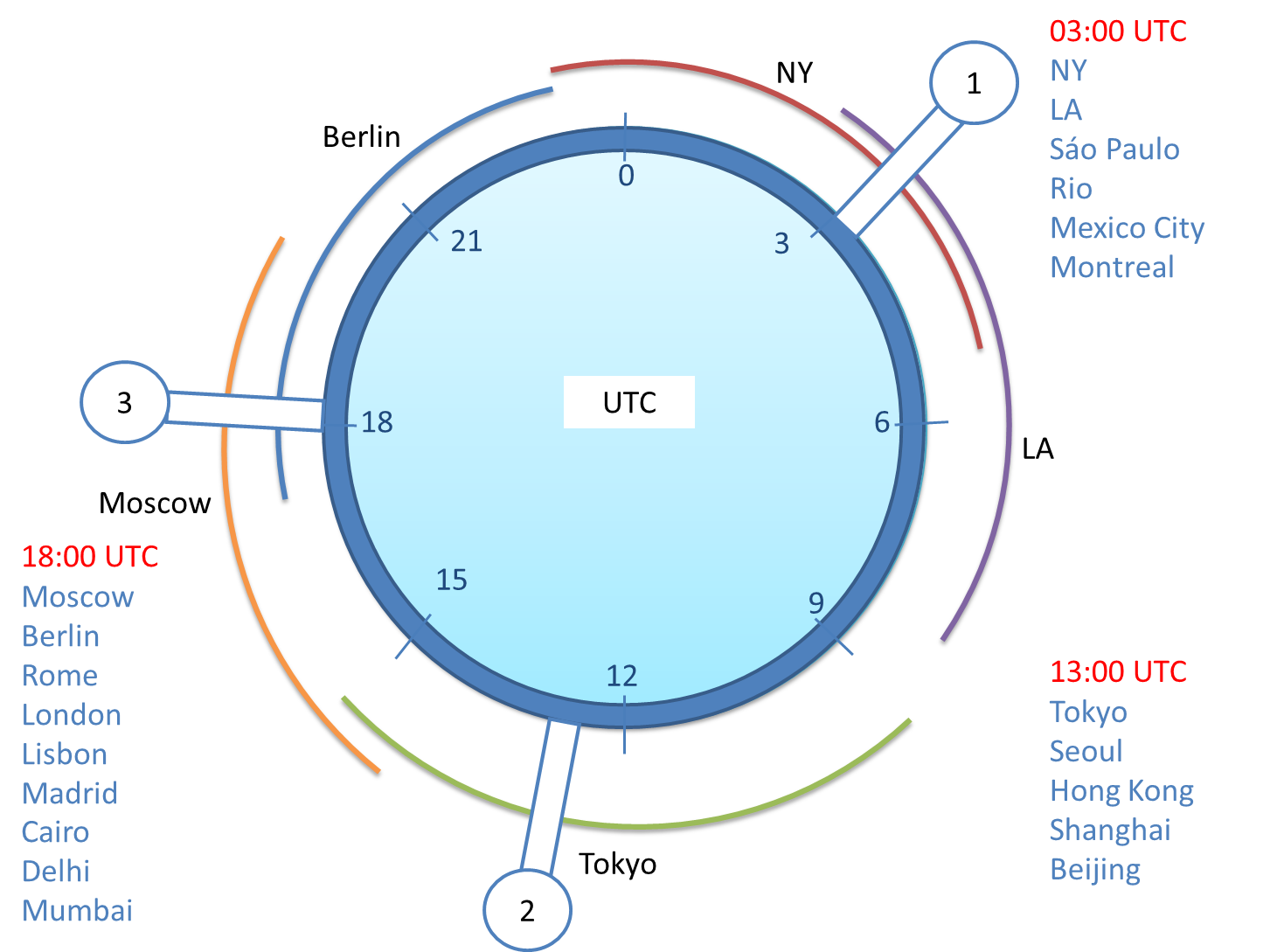}
	\caption{The dataset of 6M HTTP requests is constructed from slices of 2M each from 03:00, 13:00, and 18:00 UTC on February 2, 2012.}
	\label{fig:logsample}
\end{figure}

%\vspace{-2mm}
%%%%%%%%%%%%%%%%%%%Excluded%%%%%%%%%%%%%%%%%
%\subsection{Wayback Machine Access Logs}
%A Web server log file is a plain text file that records the activity information of the submitted requests from the users on the web server. The Wayback Machine access logs contain the following fields: client IP, access time, HTTP request method (GET or HEAD), URI, the protocol (HTTP), HTTP status code (200, 404, etc.), bytes sent, referring URI, and User-Agent. A segment from the Wayback Machine server log is shown in Figure \ref{fig:sample}. For privacy purposes, Internet Archive anonymized the client IP address. 

%\begin{figure*}
%\begin{verbatim}
%0.247.222.86 - - [02/Feb/2012:07:03:55 +0000] "GET
%http://web.archive.org/web/20020404020224/http://www.aura.vu/ 
%HTTP/1.1" 200 18875 "http://wayback.archive.org/web/*/http://www.aura.vu" 
%"Mozilla/5.0 (Macintosh; Intel Mac OS X 10_6_8) AppleWebKit/535.7 
%(KHTML, like Gecko) Chrome/16.0.912.77 Safari/535.7"}
%\end{verbatim}
%	\caption{Sample of the Wayback Machine access log.}
%	\label{fig:sample}
%\end{figure*}
%%%%%%%%%%%%%%%%%%%%%%%%%%%%%%%%%%%%%%%%%%

\section{Methodology}
We use the Internet Archive's Wayback Machine server logs in our analysis. We constructed our sample by combining three different slices of 2M records each (covering approximately 30 minutes) at times 03:00, 13:00, and 18:00 UTC on February 2, 2012, for a total dataset of 6M records. Because we are checking the language of the content accessed by the web archive users, we cover the peak time of Internet traffic periods for several countries with different language speakers to avoid biasing the results. According to many studies, the hours between 6 p.m. to 12 a.m. are considered to be peak times for Internet traffic \cite{Don2012,Fukuda2005,Todd2011}. We picked samples from the log file that were representative of the peak time for several cities around the world, as shown in Figure \ref{fig:logsample}. 
%The sample comprised 6 million requests to the Wayback Machine web server. 
%%%%%%%%%%%%%%%%%%%%%%
%\vspace{-1em}
\begin{table}
  \centering
    \begin{tabular} {|l| l| l |l |l |l| l| l|  l|}\hline 
%  & & &  \multicolumn{4}{c|}{\textbf{HTTP Status Code}} & \multicolumn{2}{c|} {\textbf{Cleaned Requests}} \\
%\cline{4-9}
GET &   Embedded & Null Ref  & 2xx & 3xx & 4xx & 5xx & Humans & Robots\\ \hline 
98.7\% & 42.9\% & 46.6\% &  33.1\% & 51.4\% & 12.0\% & 3.5\% & 1.5\% & 18.8\% \\ \hline 
\end{tabular}
  \caption{Data set statistics based on 6M requests. Note that the last two columns are the percentage of humans and robots remaining after cleaning (removing the irrelevant requests to the analysis, e.g., embedded resources).}
  \label{tab:features}
\end{table}

Table \ref{tab:features} contains the features of the sample. The features, from left to right, are the percentage of requests that used the GET method, were for embedded resources of web pages (such as images and CSS files), had null referrers (i.e., they do not identify a URI that links to a page at the Internet Archive), were successful requests (2xx status code), were redirections (3xx status code), were client errors (4xx status code), were server errors (5xx status code), remained from human requests after cleaning (removing the irrelevant requests to the analysis, e.g., embedded resources), and remained from robot requests after cleaning. The characteristics are consistent with our previous analysis of web archives \cite{AlNoamany2013}.

Preparing the Wayback access logs for usage mining starts with transforming the raw log file into server sessions through web log preprocessing (data cleaning, user identification, and session identification) \cite{Reddy2012}. A session is the group of consecutive requests performed by a user \cite{Markov2007}. We apply the same methodology as in our previous work for preprocessing the logs and web robot detection \cite{AlNoamany2013}.
\section{What do Wayback Machine Users Look for?}
In this section, we give insight into what web archive users look for in terms of the content language of requested pages. We used the language detection library created by Shuyo \cite{nakatani2010langdetect} for detecting the language.

\subsection{Archived Web Pages} 

\subsubsection{Distribution of Languages Used in the Wayback Machine}
We extracted the successful requests (HTTP 200 status code) from humans and robots to detect the language distributions for the content of the requested pages. These successful requests represent 93.1\% (85,909  out of 92,204) of all human requests and 56.7\% (639,684 out of 1,127,204) of all robots requests. The request can be for a URI-T or a URI-M. For the URI-Ts, which represent 13\% of human requested pages and 80.8\% of robot requested pages, we estimated the language by using the most recent URI-M from the TimeMap. %based on the assumption that the language will be the same for at least 100 consecutive URI-Ms of the same URI-R. 
%.
\begin{table}
\begin{tabular}{|lr|lr||lr|lr|}
\hline
\multicolumn{4}{|c||}{\textbf{URI-Ms with HTTP 200}} & \multicolumn{4}{|c|}{\textbf{URI-Rs with HTTP 404}} \\
\cline{1-8}
Language & Humans & \ \ Language & Robots & Language & Humans & \ \ Language & Robots \\ \hline \hline 
English & 71.7\% & \ \ English & 72.4\% & English & 66.9\% & \ \ English & 62.2\%  \\
Japanese & 5.5\% & \ \ Russian & 7.0\% & Russian & 7.9\% & \ \ Russian & 11.1\% \\
German & 3.6\% & \ \ German & 3.1\% & German & 5.4\% & \ \ German & 3.8\% \\
Vietnamese & 2.9\% & \ \ Spanish & 1.9\% & Japanese & 5.1\% & \ \ Indonesian & 3.1\% \\
Russian & 2.3\% & \ \ French & 1.8\% & Spanish & 2.5\% & \ \ Polish & 2.5\% \\
Portuguese & 2.1\% & \ \ Vietnamese & 1.7\% & Polish & 2.3\% & \ \ Vietnamese & 2.2\% \\
French & 2.1\% & \ \ Japanese & 1.5\% & Romanian & 1.6\% & \ \ Spanish & 2.0\% \\
Spanish & 1.9\% & \ \ Polish & 1.5\% & French & 1.2\% & \ \ Thai & 1.9\% \\
Bengali & 1.8\% & \ \ Portuguese & 1.3\% & Italian & 0.8\% & \ \ French & 1.8\% \\
Italian & 0.9\% & \ \ Thai & 1.1\% & Portuguese & 0.7\% & \ \ Dutch & 1.1\% \\ \hline 
\end{tabular}
  \caption{The top 10 languages for URI-Ms with HTTP 200 (on the left) and for the URI-Rs of unarchived requested pages (on the right).}
  \label{tab:lang}
\end{table}
%The top 10 languages  in web archives
We identified 52 different languages from the successful requests. The left two columns of Table \ref{tab:lang} show the top 10 languages which accounted for 94.8\% of human and 93.4\% of robot requests. For both human and robot users, English contributes the most to the successful requests, reflecting the high web archive penetration rate in English speaking countries. Japanese is the second most frequent language with 5.5\% for humans, but Russian is the second most frequent language for robots at 7.0\%. We also notice that despite of the existence of web archives in Europe, the requests to the IA from speakers of European languages contribute 13\% of the top 10 list for human requested pages and 18.5\% of the top 10 list for the robot requests. 

%In Section 5, we provide a detailed analysis for the referrer of human requests, investigating if there is an effect from the referrers on the distributions of the languages or not. 

\subsubsection*{Existence on the Live Web}
From all 85,909 successful human requests, we checked the existence of the 40,791 unique URI-Rs on the live web. The robots generated 639,684 successful requests, in which there are 331,573 unique URI-Rs whose existence on the live web were also checked. We also checked the pages that give ``soft 404s'', which return HTTP 200, but do not actually exist, based on the algorithm in \cite{Bar-Yossef:2004:STG:988672.988716}. Table \ref{tab:status} contains the results of checking the status of the web pages on the live web.

%p{2cm}p{2.5cm}|p{2cm}p{2.5cm}
%URIs with HTTP 200
\begin{table}
\centering
\begin{tabular}{|p{5cm}| r r|| r r|}
\cline{2-5}
\multicolumn{1}{c|}{}&\multicolumn{2}{c||}{\textbf{\ \ Found in Archive\ \ \ \ }}  & \multicolumn{2}{c|}{\textbf{\ \ \ \ Unarchived\ \ \ \ \ \ }} \\
\cline{2-5}
\multicolumn{1}{c|}{}  & Humans & Robots & Humans &Robots \\ \hline 
\textbf{URI-Rs available on live web} & 36.4\% & 62.5\% & 25.4\% & 33.2\% \\ 
\textbf{URI-Rs missing from live web} & 63.6\% & 37.5\% & 74.6\% & 66.8\% \\ \hline \hline
\textbf{Uniq. URI-Rs} & \textbf{40,791} & \textbf{331,573} & \textbf{2,441} & \textbf{209,384} \\ \hline
\end{tabular}
  \caption{The existence of the requested archived pages on the live web. Available represents the requests which ultimately return ``HTTP 200'', while missing represents the requests that return HTTP 4xx, HTTP 5xx, HTTP 3xx to others except 200, timeouts, and soft 404s.}
  \label{tab:status}
\end{table}
%\vspace{-0.5mm}

We believe humans access the Wayback Machine because they do not find web pages on the live web. Table \ref{tab:status} shows that for the requested pages that were found in the archive (returned HTTP 200 status), the percentage of the available pages on the live web for human requests is 36.4\%. On the other hand, the percentage of the available pages on the live web for robot requests is 62.5\%.  

\subsection{Unarchived Web Pages}
Of the 6M requests in our sample, 12\% returned HTTP 404 status, as shown in Table \ref{tab:features}. Not all of these are actually unarchived; approximately 2\% of the unique URI-Rs are malformed (e.g., http://http://cnn.com) and were removed. We used the remaining valid URI-Rs (209,348 robots and 2,441 humans) to detect content language, check live web status, and check existence in other archives.

%The HTTP 404 status of the requested archived pages represented 12\% of the 6M requests of our sample, Table \ref{tab:features}. First, we check the validity of the given URI-Rs for the missing pages on the archive to find that 2\% of robots and humans unique URI-Rs are malformed (e.g., http://http://cnn.com). We extracted the valid URI-Rs for robot (209,384) and human (2,441) users to detect the language of their content, check their status on the live web, and check their existence on other archives.

\subsubsection*{Existence on the Live Web}
The current state of the requested URI-Rs that had HTTP 404 status was determined by testing their existence on the live web. Of the URI-Rs that were not found in the Wayback Machine, 66.8\% of those requested by robots and 74.6\% of those requested by humans do not exist on the live web. To compensate for transient errors we repeated the requests several times for a week before declaring a URI-R non-existent.
	
\subsubsection*{Distribution of the Content-Language for Unarchived Web Pages}
We detected the content language of available URI-Rs on the live web, which represent 25.4\% (620 out of 2,441) of the unique URI-Rs for humans and 33.2\% (69,510 out of 209,384) for robots. The total number of requested URI-Rs is 227,450 for robots and 1,578 for humans. The two rightmost columns of Table \ref{tab:lang} have the results for robots and humans separately. For the web pages that were not archived in IA's Wayback Machine, English is the most requested language with 66.9\% of the human-requested web pages and 62.2\% of the robot-requested web pages. The top 10 languages compromised 94.5\% of all the content-language of the requested pages. European languages made up 22.5\% of the human-requested pages and 22.4\% of the robot-requested pages.

%European languages also contribute significantly to the content of the unarchived pages. We noticed from the results that of the 94.5\% that represent  for both human and robot requested pages, European languages contribute 22.5\% of top list of human content languages and 22.4\% of the top list of robot content languages.

%%%%%%%%%%%%%%%%%%%%%%%%%%%%%%%

\begin{table}
\centering
\begin{tabular}{ |l|l|r| r |}
  \hline
\textbf{Web Archive} & \textbf{Archive Web Site} & \textbf{\#URI-R }& \textbf{\#URI-M}  \\ \hline\hline
Internet Archive (2013) & web.archive.org & 56,503 & 1,657,264  \\ 
The National Archives & webarchive.nationalarchives.gov.uk & 787 & 15,354  \\ 
ArchiefWeb & www.archiefweb.eu &  47 & 18,347  \\
Archive-It & archive-it.org &41 & 4,682  \\ 
UK Web Archive & www.webarchive.org.uk  & 38 & 12,277  \\ 
Library of Congress & webarchive.loc.gov  & 35 & 1,092  \\
WebCite & webcitation.org & 29 & 1,104  \\ \hline
\end{tabular}
\caption{The number of the found URI-Rs and the corresponding URI-Ms of the missing pages (211,825 unique URI-Rs) on the web archives.}
\label{tab:arch}
\end{table}
%\vspace{-2em}
%%%%%%%%%%%%%%%%%%%%%%Excluded%%%%%%%%%%%%%%%%%%%%%%%%%%%%%%%55
%
%\vspace{-3mm}

\subsubsection{Existence in Other Web Archives}
We checked the 211,825 unarchived pages for existence in other archives at the time of the experiment. The existence in the web archives was tested by querying Memento proxies and aggregator \cite{memento:rfc}. For completeness and fairness, we also included the results from IA's Wayback Machine in March 2013. This resulted in 56,503 out of 211,825 URI-Rs that were unarchived in Feb. 2012 now being available in the archive. Table \ref{tab:arch} contains the number of URI-Rs found in the web archives and the number of covered URI-Ms. The Internet Archive has the most coverage at the time of experiment as they have increased their repository recently \cite{Kahle2013}.

\section{Where do Wayback Machine Users Come From?}
We used the referrer field, which contains the web page that links to the resource, for the logs in our sample to determine how people discover the Wayback Machine. In terms of sessions, 84.8\% of robot sessions do not have referrers while only 18.1\% of human sessions do not have referrers (i.e., they reached the Wayback Machine by a link in an email, direct address, or direct bookmark). An empty referral field is a strong indicator of a robot.
\begin{table}
  \centering
    \begin{tabular} {|l| r|l |}\hline 
\textbf{Web Site} & \textbf{Percentage} \ & \textbf{Description} \\ \hline \hline 
en.wikipedia.org & 12.9\%  \ & Wikipedia \\
archive.org & 11.9\% \  & IA Home Page \\
reddit.com & 10.2\% \ & Social News Web Site \\
google.TLD & 9.9\%  \ & Search Engine \\
info-poland.buffalo.edu & 1.5\%  \ & Polish Studies \\
de.wikipedia.org & 1.4\% \ & Wikipedia \\ 
cracked.com & 1.2\%  \ & Humor Site \\
snopes.com & 1.1\% \ & Urban Legends Reference Pages \\
facebook.com & 0.9\% \ & Social Media \\
crochetpatterncentral.com & 0.9\%  \ & Crocheting Hobbies \\ \hline 
\end{tabular}
  \caption{The top 10 referrers.}
  \label{tab:topref}
\end{table}

In this section, we provide a detailed analysis of the referrer field of human users to gain insight into who links to the Wayback Machine and how they link to it. Robots are not included in the analysis of referrers because the majority of robots do not have referrers and if they do, we do not necessarily trust their values.
%We plan to investigate the percentage of robots that have real referrers in upcoming study.

\subsection{Who Links to Wayback Machine?}
The percentage of human sessions with referrers is 81.9\%. We eliminated the sessions that were referred by a URI-M or URI-T because they started prior to our sample. Of the sessions that started with an external referrer, 9.6\% came from Google. The users who came from the home page of the IA contributed to 11.9\% of the sessions with referrers. That means that many people start with the IA to access the Wayback Machine.

\subsubsection{Top Referrers} Table \ref{tab:topref} contains the top 10 referrers that link to IA's Wayback Machine. The list of top 10 referrers represents 51.9\% of all the referrers. As the table shows, en.wikipedia.org outnumbers all other sites including the search engine and the home page of Internet Archive (archive.org). Note that ``google.TLD'' represents Google search and 24 other pages from Google (e.g., http://www.google.com/about/company/history.html). Since the majority are from Google search, we describe it as search engine. Facebook also appears as a top referrer, which indicates that many people share links to the past. 
\begin{table}
\centering
\begin{tabular}{|l|| l |l |l |l |l| l| l| l| l| l  |}
\hline
\textbf{TLD} & .com & .org & .net & .jp & .ru & .de & .edu & .to & .uk & .info  \\ \hline
\textbf{Percentage} & 45.4\% & 33.9\% & 8.4\% & 1.8\% & 1.4\% & 1.4\% & 1.1\% & 0.7\% & 0.6\% & 0.5\%  \\ \hline
\end{tabular}
  \caption{The top 10 TLDs of the referrers.}
  \label{tab:tld}
\end{table}
\begin{table}
\centering
\begin{tabular}{|l|| l |l |l |l |l| l| l| l| l| l  |}
\hline
\textbf{ccTLD} & .com & .uk & .de & .ca & .jp & .pl & .nl & .ru & .fr & .br  \\ \hline
\textbf{Percentage} & 56.7\% & 6.0\% & 5.3\% & 4.8\% & 3.7\% & 2.2\% & 1.9\% & 1.7\% & 1.5\% & 1.4\%  \\ \hline
\end{tabular}
  \caption{The top 10 ccTLDs of Google search referrers.}
  \label{tab:cctld}
\end{table}

%\vspace{-2em}
%\begin{figure}[H]
%	\centering
%	\includegraphics[width=1.0\textwidth, height=2.7in]{refTLD2.png}
%	\caption{The top 20 TLD of the referrers.}
%	\label{fig:tld}
%\end{figure}
%\vspace{-3em}

\subsubsection{Classification of Referrers}
Table \ref{tab:tld} presents the distribution of Top Level Domains (TLD) for the URIs that link to the IA's Wayback Machine (only the top 10 are shown). It can be noticed that most of the connections are from the .com, .org, .net, .jp, .edu, and .ru domains. Despite of the existence of many web archives in Europe, there are many European domains linking to the IA, such as .ru (Russia), .de (Germany), .fr (France), and .it (Italy). Note that .to is the TLD for a Russian language site (http://lurkmore.to/).

For the referrers from Google search, we extracted the country code top-level domain (ccTLD) of the URIs to discover the countries of the users who came to the Wayback Machine through the search engine. The results are shown in Table \ref{tab:cctld}. English-speaking countries are in the lead, followed by the European language countries.
%langrefGraphs/circos-table-oaoddyw-large20
\begin{figure}[h!]
\centering
	\includegraphics[scale=0.16]{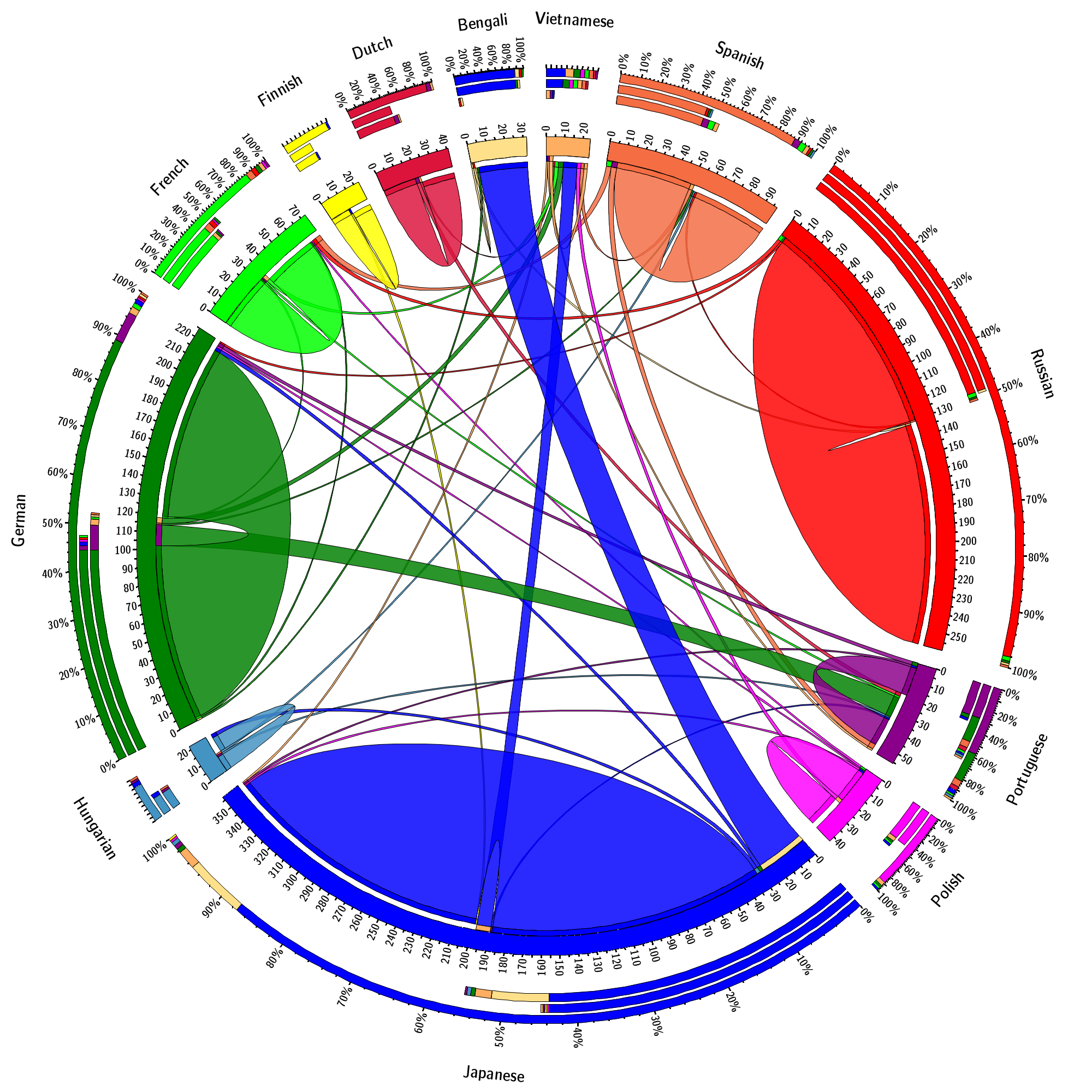}
	\caption{Most languages self-link, with the notable exceptions of Japanese $\rightarrow$ \{Bengali, Vietnamese\} and German $\rightarrow$ Portuguese.}
	\label{fig:langref}
\end{figure}

%Do the Referrers have a Significant Effect on the Distribution of the Languages of Web Archive Users?
\subsection{Inter-linking Between Languages}
From the analysis of the content languages of the referrers and the archived pages which have been linked by the referrers, English represents 80.7\% of the referrers' content languages and 80.2\% of all referred pages. English referrers link to English archived pages 92\% of the time. A small percentage of English referrers link to pages in other languages. The top 5 languages that English pages link to are (in decreasing order) Portuguese, Vietnamese, French, and German. Figure \ref{fig:langref} contains a directed weighted graph, which is created using Circos \cite{Krzywinski18062009}, for the relationship between the languages of the referrers and referees. We exclude English from the graph to be able to analyze the rest of the languages and see what they are linking to. For a particular language, the length of the outer arc represents the sum of the number of referrer pages and the number of referee pages in that language. Moving toward the center, the next arc represents the percentage of referees, and the third arc represents the percentage of referrers in that language. For example, links to Japanese archived web pages denote 46\% (160 out of 357) of all the Japanese language pages for referrers and archived pages all together. The inner circle shows the relationships between languages of the referrers and referees. Ribbons of different widths connect the languages. The direction is represented by a gap between the line and the incoming language (referrer language). For example, there are 30 links from Japanese (ja) pages to Bengali (bn) pages, which are shown as a fairly broad blue line. The languages where the relative number of referrer and referee together is less than 20 have been excluded to remove noise from the graph.

The figure shows that the languages are mainly linking to themselves with a few inter-language links. Though, recall that we have excluded English from the figure. 
Many of the top ranked languages of human-requested pages appear in the top ranked list of referrers, such as Japanese, German, Russian, Spanish, French, Polish, Dutch, Bengali, etc. It is surprising to find many European referrers to IA's Wayback Machine in spite of the existence of European web archives.

%\subsubsection{Locations of the referrers}
%Using the geolocation of the host names for the web sites of the referrer field, we detect the location of these web sites. the Map in Figure \ref{fig:maploc} shows that more than 75\% of the web sites are located in united states. 

%\begin{figure}
%	\includegraphics[width=\linewidth]{maploc.png}
%	\caption{The location of the referrers web sites that point to IA. The size %represent the in-degree and the color represents the out-degree.}
%	\label{fig:maploc}
%\end{figure}
\begin{figure}[tp]
	\centering 
	\subfigure[]{
	\includegraphics[scale=0.17]{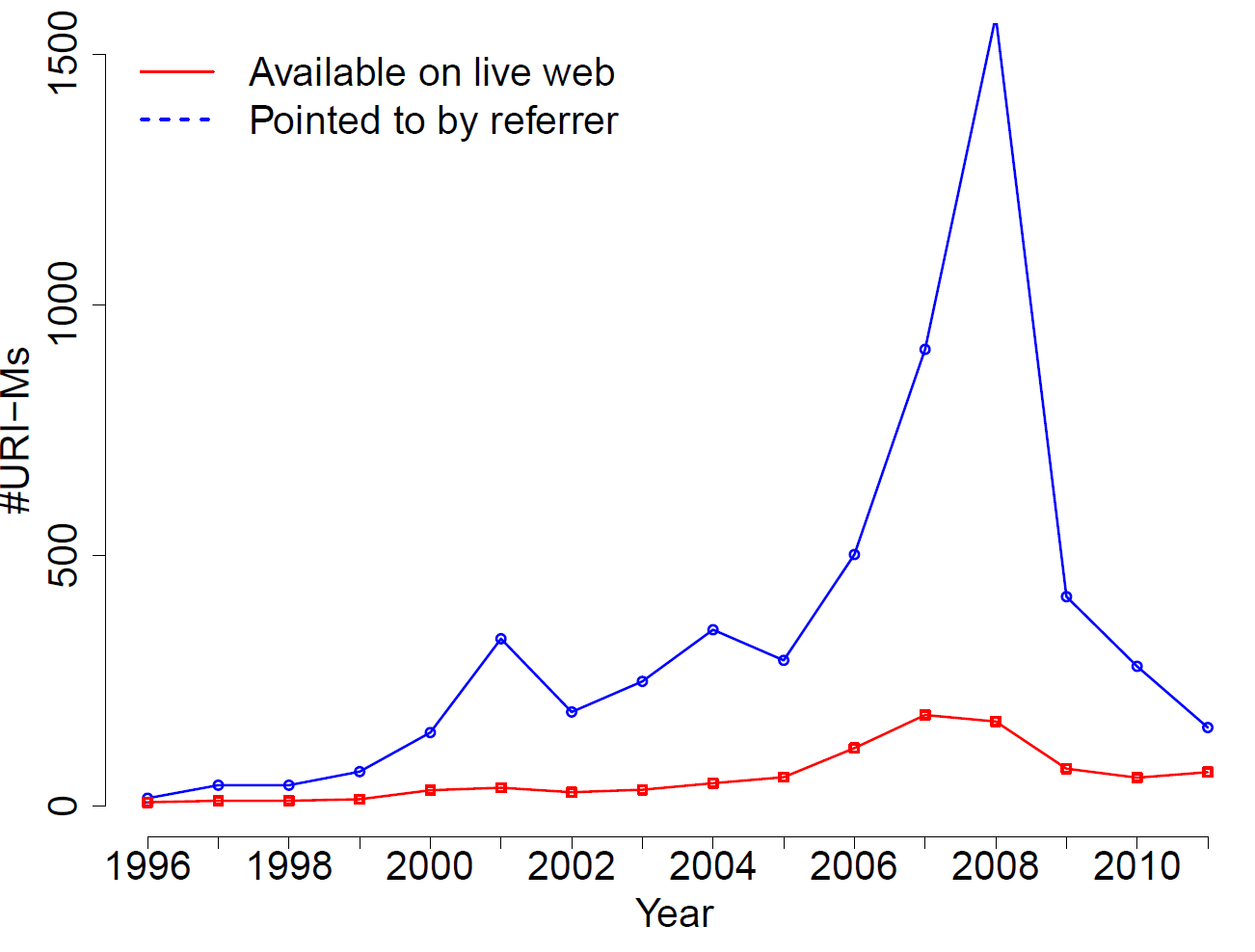}
	\label{fig:years2}
	}
	\subfigure[]{
	\includegraphics[scale=0.17]{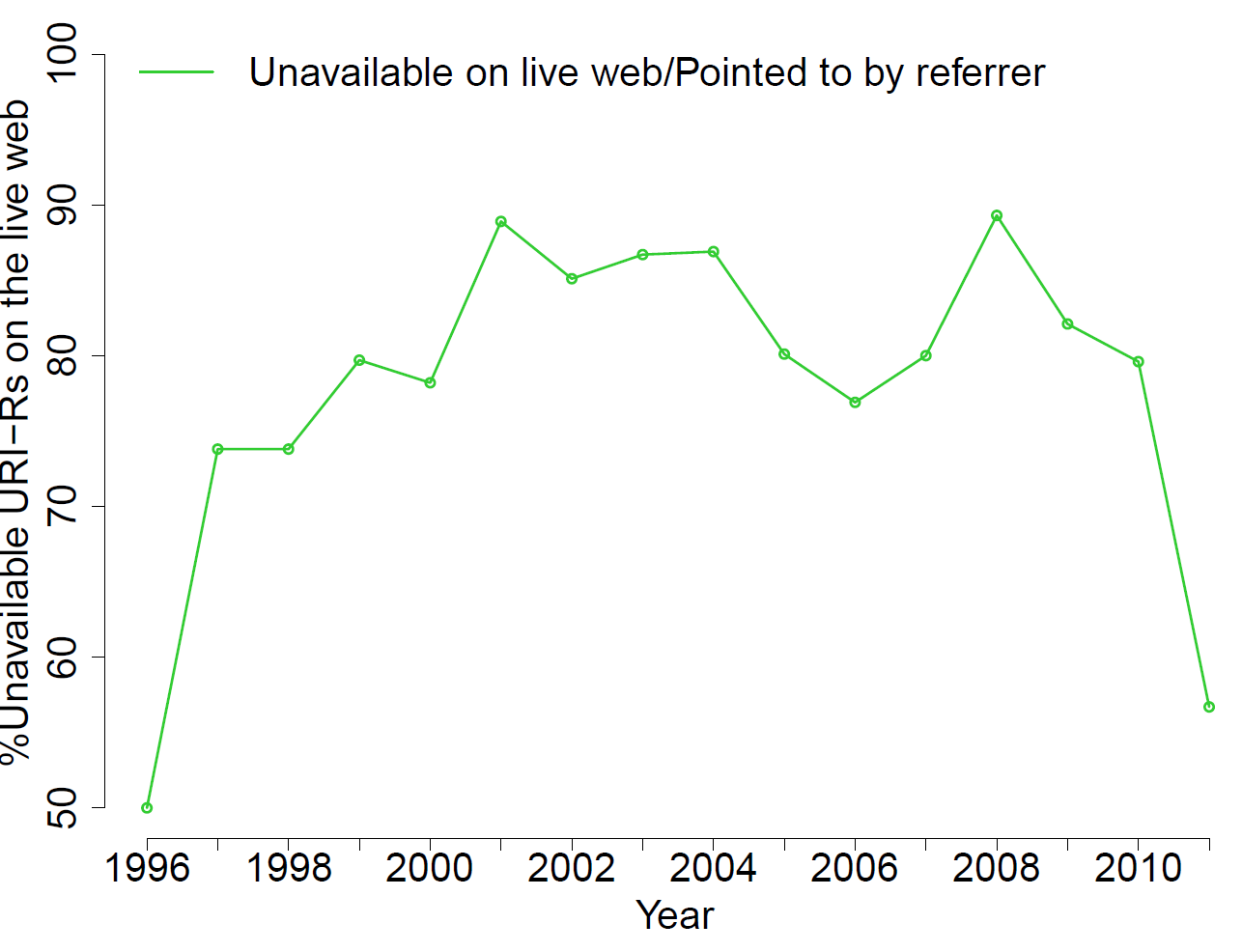}
	\label{fig:years}	
	}
	\caption{(a) The temporal distribution of URI-Ms pointed to by the referrers and the number of relative URI-Rs of these URI-Ms that are currently available on the live web. (b) The percentage of unavailable URI-Rs of these URI-Ms on the live web.}
\end{figure}

\subsection{How do Web Pages Link to the Wayback Machine?} 
We found that 86.4\% of the web pages that link to the Wayback Machine are pointing to mementos, which means they link to web pages at a specific time. There are 12.8\% of web pages that point to TimeMaps. The percentage of web pages that point to the repository (e.g., http://web.archive.org) is 0.8\%. Google search links to the top level URI, because Google does not crawl the archive based on the robots.txt exclusion protocol.

\subsubsection{Temporal Distribution of the Referred URI-Ms}
Figure \ref{fig:years2} shows the total number of mementos which were pointed to by the referrers, grouped by the year of their Memento-Datetime. There is a significant bias toward 2008, then 2007, and then a bias against the more distant past. We found 14 URI-Ms all from a single web site that link to a datetime in 2099. We assume that the referrer wants to redirect the site's visitors to the most recent copy of the linked web page. 
\subsubsection{Why do Web Sites Link to the Wayback Machine?}
The nature of the web is ephemeral, and the expected lifetime of a web pages is short \cite{Harrison:2006:JRM:1149941.1149971}. So, web archives are important to webmasters and third parties for preserving and saving many web sites. Figure \ref{fig:years} clarifies that most people link to the Wayback Machine because they did not find the pages on the live web. The figure shows that for most of the years, more than 70\% of the referred pages on the archive no longer exist on the live web.  About 83\% of all referred-to URI-Rs do not currently exist on the live web.

%	\caption{The percentage of the available URI-Rs on the live web.}

%\subsection{who stay longer?}Does the destination affect the session length?? The user who search for the archive stay longer than the users who come from external web sites. .... based on the session duration and session length average for the users who come from each category.

\section{Future Work and Conclusions}
We plan to extend our analysis for investigating if the destination of users affects the session length and the behavior of web archive users. Furthermore, we will investigate the behavior of robots in web archives more and contrast it with the behavior of robots on the live web to distinguish their behaviors.

From the analysis of Internet Archive's Wayback Machine server logs, we conclude that most humans come to the Wayback Machine to find missing pages from the live web. The percentage of the requested archived pages which currently do not exist on the live web is 65\%. We provided analysis for the distributions of languages to gain insight about what users look for. We found that English is the most used language on the Wayback Machine, followed by many European languages. European languages represent about 22\% of the web pages that were not found on the Wayback Machine, for both human and robot requests. The large percentage of European languages among the unarchived pages can be a good indicator for archival demand for European web pages. 
We also provided analysis for the human referrers to discover where Wayback Machine users come from. We discovered that wikipedia is the most frequent referrer of pages to IA's Wayback Machine. From analyzing the TLDs of the referrers, we found many European domains (.ru, .de, .fr, etc.) in the top list of the referrers. English represents 80.2\% of the referrer languages, followed by European languages. We found that the languages are linking mainly to themselves and to English. We also found that 86\% of the referrer web pages link deeply to mementos. More than 82\% of the links to these mementos are because their corresponding URI-Rs do not exist on the live web. 
%The referrers has large effect on the languages that used by humans in web archives. 

\section{Acknowledgment}
This work was supported in part by the NSF (IIS 1009392) and the Library of Congress. We thank Kris Carpenter Negulescu (Internet Archive) for access to the anonymized Wayback Machine logs.

\bibliographystyle{splncs_srt}
\bibliography{TPDL-AlNoamany}

\begin{thebibliography}{10}

\bibitem{AlNoamany2013}
AlNoamany, Y., Weigle, M.C., Nelson, M.L.:
\newblock {Access Patterns for Robots and Humans in Web Archives}.
\newblock In: Proceedings of the 13th ACM/IEEE-CS Joint Conference on Digital
  Libraries. JCDL '13 (July 2013)

\bibitem{Bar-Yossef:2004:STG:988672.988716}
Bar-Yossef, Z., Broder, A.Z., Kumar, R., Tomkins, A.:
\newblock Sic transit gloria telae: towards an understanding of the web's
  decay.
\newblock In: Proceedings of the 13th International Conference on World Wide
  Web. WWW '04, ACM (2004)  328--337

\bibitem{Carmel2008}
Carmel, D., Yom-Tov, E., Roitman, H.:
\newblock Enhancing digital libraries using missing content analysis.
\newblock In: Proceedings of the 8th ACM/IEEE-CS Joint Conference on Digital
  Libraries. JCDL '08, ACM (2008)  1--10

\bibitem{Costa2011}
Costa, M., {J. Silva}, M.:
\newblock {Characterizing Search Behavior in Web Archives}.
\newblock In: Proceedings of Temporal Web Analytics Workshop. TWAW (2011)

\bibitem{costa2010}
Costa, M., Silva, M.J.:
\newblock {Understanding the Information Needs of Web Archive Users}.
\newblock In: Proc. of the 10th International Web Archiving Workshop. (Sept
  2010)

\bibitem{Fukuda2005}
Fukuda, K., Cho, K., Esaki, H.:
\newblock {The impact of residential broadband traffic on Japanese ISP
  backbones}.
\newblock SIGCOMM Comput. Commun. Rev. \textbf{35}(1) (January 2005)

\bibitem{Harrison:2006:JRM:1149941.1149971}
Harrison, T.L., Nelson, M.L.:
\newblock {Just-In-Time Recovery of Missing Web Pages}.
\newblock In: Proceedings of the 17th Conference on Hypertext and Hypermedia.
  HYPERTEXT '06, ACM (2006)  145--156

\bibitem{Kahle2013}
Kahle, B.:
\newblock {Wayback Machine: Now with 240,000,000,000 URLs}.
\newblock \url{http://blog.archive.org/2013/01/09/updated-wayback/} (January
  2013)

\bibitem{Krzywinski18062009}
Krzywinski, M.I., Schein, J.E., Birol, I., Connors, J., Gascoyne, R., Horsman,
  D., Jones, S.J., Marra, M.A.:
\newblock Circos: An information aesthetic for comparative genomics.
\newblock Genome Research (2009)

\bibitem{Markov2007}
Markov, Z., Larose, D.T.:
\newblock {Data Mining the Web: Uncovering Patterns in Web Content, Structure,
  and Usage}.
\newblock John Wiley \& Sons, Inc. (2007)

\bibitem{wayback:billion}
Negulescu, K.C.:
\newblock {Web Archiving @ the Internet Archive}.
\newblock Presentation at the 2010 Digital Preservation Partners Meeting,
  \url{http://1.usa.gov/XSjDG8} (2010)

\bibitem{Padia2012}
Padia, K., AlNoamany, Y., Weigle, M.C.:
\newblock {Visualizing Digital Collections at Archive-It}.
\newblock In: Proceedings of the 12th ACM/IEEE-CS Joint Conference on Digital
  Libraries. JCDL '12, ACM (2012)  15--18

\bibitem{Reddy2012}
Reddy, K.S., Varma, G.P.S., Babu, I.R.:
\newblock {Preprocessing the Web Server Logs – An illustrative approach for
  effective usage mining}.
\newblock ACM SIGSOFT Software Engineering Notes \textbf{37}(3) (May 2012)
  1--5

\bibitem{Don2012}
Reisinger, D.:
\newblock {Netflix gobbles a third of peak Internet traffic in North America}.
\newblock CNET, \url{http://goo.gl/2cVPg} (2012)

\bibitem{nakatani2010langdetect}
Shuyo, N.:
\newblock {Language Detection Library for Java}.
\newblock \url{http://code.google.com/p/language-detection/} (2012)

\bibitem{Silva2009}
Silva, A.J.C., Gon\c{c}alves, M.A., Laender, A.H.F., Modesto, M.A.B., Cristo,
  M., Ziviani, N.:
\newblock Finding what is missing from a digital library: A case study in the
  computer science field.
\newblock Inf. Process. Manage. \textbf{45}(3) (May 2009)  380--391

\bibitem{Thelwall2004}
Thelwall, M., Vaughan, L.:
\newblock A fair history of the web? examining country balance in the internet
  archive.
\newblock Library \& Information Science Research \textbf{26}(2) (2004)

\bibitem{Tofel2007}
Tofel, B.:
\newblock {Wayback for Accessing Web Archives}.
\newblock In: Proceedings of International Web Archiving Workshop. IWAW (2007)

\bibitem{memento:rfc}
{Van de Sompel}, H., Nelson, M.L., Sanderson, R.:
\newblock {HTTP framework for time-based access to resource states -- Memento}.
\newblock \url{https://datatracker.ietf.org/doc/draft-vandesompel-memento/}
  (2012)

\bibitem{nelson:memento:tr}
{Van de Sompel}, H., Nelson, M.L., Sanderson, R., Balakireva, L.L., Ainsworth,
  S., Shankar, H.:
\newblock {Memento: Time Travel for the Web}.
\newblock Technical Report arXiv:0911.1112 (2009)

\bibitem{Todd2011}
Wasserman, T.:
\newblock {Netflix takes up 32.7\% of Internet bandwidth}.
\newblock Marshable, \url{http://goo.gl/2FtWa} (2011)

\bibitem{Zhuang2005}
Zhuang, Z., Wagle, R., Giles, C.:
\newblock What's there and what's not?: focused crawling for missing documents
  in digital libraries.
\newblock In: Proceedings of the 5th ACM/IEEE-CS Joint Conference on Digital
  Libraries. JCDL '05 (2005)  301--310

\end{thebibliography}

\end{document}